\documentclass[iop]{emulateapj}

\usepackage{color}
\usepackage{graphicx}
\usepackage{amsmath}

\shorttitle{Testing $f(R)$ gravity with GW date from Einstein
Telescope} \shortauthors{Pan et al.}

\newcommand{\beq}{\begin{equation}}
\newcommand{\eeq}{\end{equation}}

\def\ba{\begin{eqnarray}}
\def\ea{\end{eqnarray}}

\begin{document}

\title{Testing $f(R)$ gravity with the simulated data of gravitational waves from the Einstein Telescope}

\author{Yu Pan$^{1}$, Yuan He$^{1}$, Jing-Zhao Qi$^{2\ast}$, Jin Li$^{3}$, Shuo Cao$^{4\dag}$, Tonghua Liu$^{4}$, and Jun Wang $^{5}$}
\affiliation{1. School of Science, Chongqing University of Posts and Telecommunications, Chongqing 400065, China; \\
2. Department of Physics, College of Sciences, Northeastern University, Shenyang 110004, China; \emph{qijingzhao@mail.neu.edu.cn} \\
3. Department of Physics, Chongqing University, Chongqing 400030, China; \\
4. Department of Astronomy, Beijing Normal University, Beijing 199875, China; \emph{caoshuo@bnu.edu.cn} \\
5. School of Physics and Astronomy, Yunnan University, Kunming 650091, China}

%

\begin{abstract}

In this paper we analyze the implications of gravitational waves
(GWs) as standard sirens on the modified gravity models by using the
third-generation gravitational wave detector, i.e., the Einstein
Telescope. Two viable models in $f(R)$ theories within the Palatini
formalism are considered in our analysis
($f_{1}(\mathcal{R})=\mathcal{R}-\frac{\beta}{\mathcal{R}^{n}}$ and
$f_{2}(\mathcal{R})=\mathcal{R}+\alpha\ln{\mathcal{R}}-\beta$), with
the combination of simulated GW data and the latest electromagnetic
(EM) observational data (including the recently released Pantheon
type Ia supernovae sample, the cosmic chronometer data, and
baryon acoustic oscillation distance measurements). Our analysis
reveals that the standard sirens GWs, which provide an independent
and complementary alternative to current experiments, could
effectively eliminate the degeneracies among parameters in the two
modified gravity models. In addition, we thoroughly investigate the
nature of geometrical dark energy in the modified gravity theories
with the assistance of $Om(z)$ and statefinder diagnostic analysis.
The present analysis makes it clear-cut that the simplest
cosmological constant model is still the most preferred by the
current data. However, the combination of future naturally improved
GW data most recent EM observations will reveal the consistency or
acknowledge the tension between the $\Lambda$CDM model and modified
gravity theories.

\end{abstract}

\section{Introduction}

The measurement of cosmic acceleration is crucial for our
understanding of the nature of the Universe. A number of scenarios
have been proposed to explain this remarkable discovery, which fall
into two general categories. In the framework of one approach, an
exotic component called dark energy is introduced to act the driving
force behind the cosmic acceleration \citep{Cao11,Cao14}. Until now,
all of the independent astrophysical observations are well
consistent with the existence of vacuum energy density or a
non-vanishing cosmological constant $\Lambda$
\citep{Cao12a,Cao12b,Cao12c,Pan15,Cao15a,Cao15b,Cao17a,Cao17b,Li17,Risaliti18,Liu19,Ma19}.
However, such simplest cosmological constant model is still
embarrassed by the coincidence problem and fine-tuning problem,
which triggered great efforts to understand the evolution of the
Universe at early time \citep{Abbott17,Qi19a,Cao19a,Cao20,Zheng20}.
On the other hand, such an accelerated expansion is possible with
the modification of the gravitational sector in Einstein-Hilbert
action \citep{Bao07,Carroll05,Capozziello05,Felice10,Sotiriou10}.
Following this direction, a lot of extended theories of gravity have
been proposed in the past decades, i.e., Gauss-Bonnet gravity
\citep{Nojiri05}, $f(T)$ gravity \citep{Bengochea09,Cai15,Qi16}, and
$f(R)$ gravity \citep{Felice10,Sotiriou10,Wang18}. Especially, the
$f(R)$ theories of gravity (in which $f(R)$ is a function of the
Ricci scalar $R$) are widely employed in modern cosmology, benefit
from its advantages of describing the large scale structure
distribution of the universe \citep{Dos16,Voivodic17,Wang13} and
explaining the late-time cosmic acceleration. Note that
there are mainly two different variational approaches providing
various interesting mechanisms, i.e., the metric formalism and the
Palatini formalism \citep{Bao07,Santos08,Capozziello08}. In the
former formalism, by assuming the connections to be the Christoffel
symbols defined in terms of the metric, one can get the generalized
field equations by varying the action with respect to the metric,
which leads to the fourth-order equations difficult to deal with in
practice. Furthermore, it was found in the recent studies that some
models within such framework may not only exhibit violent
instabilities in the weak gravity regime of matter
\citep{Dolgov03,Nojiri03,Wang10}, but also fail to produce a
standard matter-dominated era followed by an accelerated expansion
\citep{Amendola06a,Amendola06b}. In this paper, we will mainly
investigate the constraints on the Palatini formalism model
($R=g^{\mu\nu}R_{\mu\nu}(\Gamma)$), in the framework of which the
metric and connection are treated as independent dynamical variables
in the action \citep{Vollick03,Felice10,Sotiriou10}. The advantage
of the Palatini approach is that the resultant equations are of
second order, which are different from the ones derived in the
metric approach but more concordant with the field equations in
other branches of physics \citep{Vollick03,Felice10,Sotiriou10}.
Specifically, Palatini gravity have been studied in the context of
dark energy models, with the late-time constraints from various
astrophysical observations
\citep{Bekov20,Baghram20,Aoki20,Leanizbarrutia17,Bohmer13,Capozziello13,Harko12,Gu18,Rosa20,Rosa17,Borowiec12}.
Meanwhile, it is interesting to note that some models could exhibit
very promising behaviors, i.e., they not only correctly fit
different observational data, but also well reproduce the
cosmological evolution with early and late-time acceleration
\citep{Yang09,Fay07,Seokcheon07}. We refer to \citet{Sotiriou10} for
a review on the $f(R)$ gravity.

From observational point of view, it is important to investigate
whether these modified gravity theories are indeed compatible with
different kinds of observational data. Currently, such analysis has
been performed in the EM domain, focusing on different types of
standard cosmological probes: type Ia supernovae (SNe Ia), baryon
acoustic oscillations (BAO), cosmic microwave background (CMB), and
intermediate-luminosity radio quasars (ILQSO)
\citep{Amarzguioui06,Fay07,de16,Song07,Xu18,Santos08}. For instance,
the first attempt to derive constraints on the $f(R)$ model
$f(R)=R-\frac{\beta}{R^n}$ was presented in \citet{Amarzguioui06},
which was then extended in \citep{Fay07} by employing the first year
Super-Nova Legacy Survey (SNLS) data, the baryon acoustic
oscillation peak in the SDSS luminous red galaxy sample and the CMB
shift parameter. It was found that in subsequent analysis that such
$f(R)$ model could produce the sequence of radiation-dominated,
matter-dominated, and accelerating periods without the inclusion of
dark energy \citep{Santos08}. However, it is necessary and important
to have other complementary cosmological probes in the GW domain,
especially the inspiraling and merging compact binaries consisting
of neutron stars (NSs) and black holes (BHs) \citep{Schutz86}. It is
well known that the first direct detection of gravitational waves
(GWs) by the LIGO/Virgo collaboration \citep{Abbott16} has opened
the era of GW astronomy. As a promising high-redshift complementary
tool to SNe Ia, the greatest advantage of GWs lies in the fact that
the distance calibration of such standard siren is independent of
any other distance ladders \citep{Taylor12}. Using the waveform
signal to directly measure the luminosity distance $D_L$ to the GW
sources, the possible cosmological application of these standard
sirens has been extensively discussed in the literature
\citep{Holz05,Dalal06,Zhao11,Cai17,Weijunjie19,Qi19,Liao19,Cao19b,Wu19,Liu20}.
More interestingly, it was originally proposed
\citep{Rocco19} that standard sirens from binary neutron star
mergers could provide strong observational bounds on modified
gravity theories (scalar-tensor theories). Specially, the final
results showed that the $f(R)$ gravity (i.e., the Hu-Sawicki $f(R)$
gravity model) can be tested at very high precision (95\% confidence
level), based on the third generation ground-based GW detector
(Einstein telescope). Such conclusion was further supported by
recent measurements of Hubble constant in the context of modified
gravity theories ($f(R)$ and $f(T)$ models), based on the combined
geometrical EM data sets obtained in a model-independent way
\citep{Rocco20}. In this paper, we will mainly investigate the
constraints on two viable models in $f(R)$ theories within the
Palatini formalism
($f_{1}(\mathcal{R})=\mathcal{R}-\frac{\beta}{\mathcal{R}^{n}}$ and
$f_{2}(\mathcal{R})=\mathcal{R}+\alpha\ln{\mathcal{R}}-\beta$),
focusing on an updated sample of GW events based on the
third-generation GW ground-based detector, Einstein Telescope (ET).
The latest electromagnetic (EM) observational data, including the
recently released Pantheon type Ia supernovae sample, the cosmic
chronometers sample and the recent baryon acoustic oscillation
measurements are also included in our analysis, in order to achieve
a reasonable and compelling constraints on the modified gravity
models in both the electromagnetic (EM) and gravitational wave (GW)
window.

This paper is organized as follows. In Sec.~2, we review the $f(R)$
theories of gravity, mainly with two variable models within the
Palatini formalism. The observational data (GW+EM) and the
corresponding constraint results are introduced in Sec.~3 and 4,
respectively. In Sec.~5, we present some model diagnostics which
results in distinction between $f(R)$ models and $\Lambda$CDM.
Finally, we discuss our results and present our main conclusions in
Sec.~6.

\section{The $f(R)$ theory in Palatini formalism}\label{theory}

We briefly summarize the basic features of $f(R)$ theories of
gravity in the Palatini formalism \citep{Sotiriou10,Will81}. The
modified Einstein-Hilbert action is given by
\begin{equation}\label{actionJF}
S = \frac{1}{16\pi G}\int d^4x\sqrt{-g}f(\mathcal{R}) + S_m\,,
\end{equation}
where $G$ is the gravitational constant, $g$ is the determinant of
the metric tensor, and $S_m$ is the standard action for the matter
fields. In the Palatini formalism, the metric and the connection are
completely independent variables. Therefore, the definition of the
Ricci scalar can be expressed as
$\mathcal{R}=g^{\alpha\beta}\mathcal{R}_{\alpha\beta}(\tilde\Gamma_{\mu\nu}^{\rho})$,
where $\tilde\Gamma_{\mu\nu}^{\rho}$ is the affine connection, which
is different from the Levi-Civita connection
$\Gamma_{\mu\nu}^{\rho}$ in the metric formalism.

Varying the action with respect to the metric and the connection,
the field equations can be obtained as
\begin{equation}\label{field_eq}
f'\mathcal{R}_{(\mu\nu)} - \frac{f}{2}g_{\mu\nu}  = 8{\pi}GT_{\mu\nu}\,,
\end{equation}
\begin{equation}\label{field_eq c}
\tilde{\nabla}_\alpha[f^\prime(\mathcal{R})\sqrt{-g}g^{\mu\nu}] = 0\,,
\end{equation}
with $f'\equiv df/d\mathcal{R}$ and $f''\equiv d^2f/d\mathcal{R}^2$.
Here $T_{\mu\nu}\equiv -\frac{2}{\sqrt{-g}}\frac{\delta
S_{m}}{\delta g^{\mu\nu}}$ denotes the energy-momentum tensor of
matter, and $\tilde{\nabla}_{\alpha}$ represents the covariant
derivative corresponding to the affine connection
$\tilde\Gamma_{\mu\nu}^{\rho}$. It is obvious that
Eq.~(\ref{field_eq}) will reduce to the Einstein's field equations
when $f(\mathcal{R})=\mathcal{R}$, while the Einstein's field
equations in $\Lambda$CDM model can be reproduced when
$f(\mathcal{R})=\mathcal{R}-2\Lambda$.

By introducing a metric conformal $h_{\mu\nu} =
f^\prime(\mathcal{R})g_{\mu\nu}$ to Eq.~(\ref{field_eq c}), one
could derive the definition of the usual Levi-Civita connection in
terms of the new metric $h_{\mu\nu}$:
\begin{equation}
\tilde{\Gamma}^\lambda _{\ \mu\nu} =
\frac{h^{\lambda\sigma}}{2}(h_{\nu\sigma,\mu} + h_{\mu\sigma,\nu} -
h_{\mu\nu,\sigma}).
\end{equation}
Then the relation between the generalized affine connection
$\tilde{\Gamma}^\lambda _{\ \mu\nu}$ and the metric $g_{\mu\nu}$ can
be obtained as
\begin{equation}
\tilde{\Gamma}^\lambda _{\ \mu\nu} = \Gamma^\lambda _{\ \mu\nu} +
\frac{1}{2f^\prime}[2\delta^\lambda _{\ (\mu}\partial _{\nu
)}f^\prime - g^{\lambda\tau}g_{\mu\nu}\partial _\tau f^\prime].
\end{equation}
Now the relationship between the generalized Ricci tensor
$\mathcal{R}_{\mu\nu}$ and the original Ricci tensor $R_{\mu\nu}$ is
given by
\begin{equation}
\mathcal{R}_{\mu\nu} = R_{\mu\nu} - \frac{3}{2} \frac{\nabla_\mu
f^\prime \nabla_\nu f^\prime}{f^{\prime 2}} +
\frac{\nabla_\mu\nabla_\nu f^\prime}{f^\prime} +
\frac{1}{2}g_{\mu\nu}\frac{\nabla^\mu\nabla_\nu f^\prime}{f^\prime},
\end{equation}
where
\begin{equation} \label{general_Ric}
 \mathcal{R}_{\mu\nu} \equiv \tilde\Gamma^{\alpha}_{\mu\nu,\alpha}-\tilde\Gamma^{\alpha}_{\mu\alpha,\nu} +
\tilde\Gamma^{\alpha}_{\alpha\lambda}\tilde\Gamma^{\lambda}_{\mu\nu} -
\tilde\Gamma^{\alpha}_{\mu\lambda}\tilde\Gamma^{\lambda}_{\alpha\nu}\,.
\end{equation}

In the framework of the flat Friedman-Robertson-Walker (FRW) metric
\begin{equation}
\mathrm{d}s^2 = -\mathrm{d}t^2 + a^2 (t)(\mathrm{d}x^2 +
\mathrm{d}y^2 + \mathrm{d}z^2 )\,
\end{equation}
we take the energy-momentum tensor of matter as a perfect fluid
\begin{equation}\label{T}
T_{\mu\nu} = (\rho+p)U_{\mu}U_{\nu}+pg_{\mu\nu},
\end{equation}
where $\rho$ is the energy density, $p$ is the pressure and the
four-velocity $U_{\mu}$ satisfies $U_{\mu}U^{\mu} = -1$ and
$U^{\mu}U_{\mu;\nu} = 0$. The Friedmann equation of the $f(R)$ model
can be written as
\begin{equation}\label{Friedman_eq}
6(H + \frac{1}{2}
\frac{\dot{f}^\prime}{f^\prime})^2 = \frac{ 3f - \mathcal{R}f^\prime
}{f^\prime},
\end{equation}
where $H\equiv \dot{a}/a$ is the Hubble parameter, and the overdot
denotes the derivative with respect to the cosmic time $t$. Based on
the trace of Eq.~(\ref{field_eq}), i.e.,
$\mathcal{R}f^\prime(\mathcal{R})-2f(\mathcal{R}) = \kappa T$ and
the conservation equation, i.e., $\dot{\rho}_{\rm m} +3H\rho_{\rm
m}=0$, we obtain
\begin{equation}\label{R_dot}
\dot{\mathcal{R}} = \frac{3H\kappa\rho_{\rm
m}}{\mathcal{R}f^{\prime\prime}-f^\prime}.
\end{equation}
Substituting Eq.~(\ref{R_dot}) into Eq.~(\ref{Friedman_eq}), one can
numerically compute the Friedman equation as
\begin{equation}
\label{Friedman_eq1} H^2 = \frac{ 3f - \mathcal{R}f^\prime }{6f^\prime\eta^2},
\end{equation}
where
\begin{equation}
\eta = 1-
\frac{3}{2}\frac{f^{\prime\prime}}{f^\prime}\frac{\mathcal{R}f^\prime
-2f}{\mathcal{R}f^{\prime\prime}-f^\prime}.
\end{equation}
According to the definition of the angular diameter distance at
redshift $z$, one also has
\begin{eqnarray}
\label{D_A} D_{\rm A}(z)&=&\frac{1}{1+z}\int _0 ^z
\frac{\mathrm{d}z}{H(z)}
\\ \nonumber
&=&\frac{1}{3}(2f-\mathcal{R}f^\prime)^{-\frac{1}{3}}\int^{\mathcal{R}_z}_{\mathcal{R}_0}{\frac{\mathcal{R}f^{\prime\prime}-f^\prime}
{(2f-\mathcal{R}f^\prime)^\frac{2}{3}}\frac{d\mathcal{R}}{H(\mathcal{R})}}.
\end{eqnarray}

Following the procedure proposed in \citet{Fay07}, the generalized
Friedmann equation can be written in terms of redshift $z=a_0/a -1$
and the matter density parameter $\Omega_{m} \equiv
\kappa\rho_{m0}/(3H_0^2)$ as
\begin{equation}
\label{fe3}
\frac{H^2}{H_0^2} = \frac{3\Omega_{m0}(1 + z)^3 + f/H_0^2}{6f'\xi^2}\,,
\end{equation}
where
\begin{equation}
\label{xi}
 \xi = 1 + \frac{9}{2}\,\frac{f''}{f'}\,\frac{H_0^2\Omega_{m0}(1+z)^3}{Rf'' -
 f'}.
\end{equation}
The trace of Eq.~(\ref{field_eq}) leads to
\begin{equation}
\label{trace2}
\mathcal{R}f' - 2f = -3H_0^2\Omega_{m0}(1 + z)^3\,,
\end{equation}
based on which one may find that Eq.~(\ref{fe3}) will reduce to
General Relativity for the Einstein-Hilbert Lagrangean case
($f(\mathcal{R})=\mathcal{R}$). Moreover, it can be clearly seen
that by taking $f(\mathcal{R}) = \mathcal{R} -
\beta/\mathcal{R}^{n}$, Eq.~(\ref{trace2}) evaluated at $z=0$
imposes the following relation
\begin{equation}
\beta = \frac{\mathcal{R}_0^{n+1}}{n+2}\,\left( 1 - \frac{3\Omega_{m}H^2_0}{\mathcal{R}_0} \right)\,,
\end{equation}
where the present value of the Ricci scalar $\mathcal{R}_0$ is
determined from the algebraic equation resulting from
Eq.~(\ref{trace2}) and (\ref{fe3}) for $z=0$. Therefore, in the
framework of $f(R)$ parametrization in the form of
$f(\mathcal{R})=\mathcal{R}-\beta/\mathcal{R}^{n}$ and
$f(\mathcal{R})=\mathcal{R}+\alpha{ln\mathcal{R}}-\beta$
\citep{Fay07}, the constraint results on the $f(R)$ model can be
shown in the($n,\Omega_{m}$) and ($\alpha,\Omega_{m}$) plane.

Now it is necessary to comment on the theoretical
applicability of the parameterizations used in this paper. For the
former form of $f(\mathcal{R})=\mathcal{R}-\beta/\mathcal{R}^{n}$,
\citet{Capozziello03,Capozziello04} have extensively discussed the
possibility of using the models in the metric approach to generate a
late-time acceleration of the Universe. Their results showed that
for $n>0$ the metric approach was unable to give rise to a standard
matter-dominated era followed by a cosmic acceleration
\citep{Amendola06a}. However, such theories using the Palatini
approach could provide an effective solution to the above
difficulties, as was proved in the subsequent analysis of
\citep{Fay07}. Similarly, for the latter form of
$f(\mathcal{R})=\mathcal{R}+\alpha{ln\mathcal{R}}-\beta$, the
numerical results have also proved the capability of the Palatini
approach to produce the sequence of radiation-dominated,
matter-dominated, and de Sitter periods \citep{Fay07}. Therefore, in
this paper we shall focus on a variable $f(R)$ theories recently put
forward within the Palatini formalism, the cosmological dynamics of
which is still not fully understood up to now. Despite these
advantages, some concerns have been expressed on several problems of
Palatini $f(R)$ gravity, due to the differential structure of its
field equations \citep{Sotiriou10}, as well as its possible
difficulty in passing the solar system tests and providing the
correct Newtonian limit \citep{Flanagan04a,Flanagan04b}. From
theoretical point of view, such issue could be potentially fixed by
the inclusion of extra terms quadratic in the Ricci and/or Riemann
tensor, which is hard to be rigorously accounted in context of
cosmological studies like in this paper.

\section{GW simulations and EM observations} \label{obs}

In this subsection we briefly introduce the method of simulating GW
events from Einstein Telescope. Compared with the current advanced
ground-based detectors (i.e., the advanced LIGO and Virgo
detectors), such third-generation GW detector composed of three
collocated underground detectors is designed to be ten times more
sensitive, especially in the frequency range of $1-10^4$ Hz
\footnote{The Einstein Telescope Project, https://www.et-gw.eu/et/}.

We focus on the GW signals generated by the coalescence of binary
systems with component masses $m_1$ and $m_2$, defining the total
mass $M=m_1+m_2$, the symmetric mass ratio $\eta=m_1 m_2/M^2$, and
the chirp mass $\mathcal{M}_c=M \eta^{3/5}$. From observational
point of view, the observed chirp mass in the observer frame can be
written as $\mathcal{M}_{c,\rm obs}=(1+z)\mathcal{M}_{c,\rm phys}$.
Note that the GW amplitude depends on the so-called chirp mass and
the luminosity distance $D_L$, while the former can be measured from
the GW signal's phasing. Therefore, the chirping GW signals from
inspiraling compact binary stars (NS and BH) can provide an absolute
measure of the luminosity distance. Following the previous analysis
of \citet{Zhao11,Sathyaprakash10}, the mass of each black hole and
neutron star is assumed to be uniformly distributed in the range
[3,10] $M_{\bigodot}$ and [1,2] $M_{\bigodot}$. The ratio between
NS-NS and BH-NS systems is taken to be $0.03$, as is predicted by
the Advanced LIGO-Virgo network \citep{Abadie10}.

GW detectors based on the ET could measure the strain (or the time
domain waveform $h(t)$) of GWs, the Fourier transform
$\mathcal{H}(f)$ of which could be derived by applying the
stationary phase approximation
\begin{align}
\mathcal{H}(f)=\mathcal{A}f^{-7/6}\exp[i(2\pi
ft_0-\pi/4+2\psi(f/2)-\varphi_{(2.0)})], \label{equa:hf}
\end{align}
where the Fourier amplitude $\mathcal{A}$ is given by
\begin{align}
\mathcal{A}=&~~\frac{1}{D_L}\sqrt{F_+^2(1+\cos^2(\iota))^2+4F_\times^2\cos^2(\iota)}\nonumber\\
            &~~\times \sqrt{5\pi/96}\pi^{-7/6}\mathcal{M}_c^{5/6},
\label{equa:A}
\end{align}
See \citet{Zhao11} for the definition of the epoch of the merger
$t_0$, the functions of $\psi$ and $\varphi_{(2.0)}$, the angle of
inclination of the binary's orbital angular momentum $\iota$, and
the two interferometers' antenna pattern functions ($F_{+}$,
$F_{\times}$) that depend on the position of GW source, as well as
the location and orientation of the detector.

The uncertainty of luminosity distance extracted from the GW signals
can be divided into two parts. The first source of uncertainty is
quantified by the instrumental uncertainty with the form of
\begin{align}
\sigma_{D_L}^{\rm inst}\simeq \sqrt{\left\langle\frac{\partial
\mathcal H}{\partial D_L},\frac{\partial \mathcal H}{\partial
D_L}\right\rangle^{-1}},
\end{align}
Assuming that this parameter is uncorrelated with any other GW
parameters, we add a factor of 2 in the instrumental error to take
the effect of inclination into consideration \citep{Li15}
\begin{align}
\sigma_{D_L}^{\rm inst}\simeq \frac{2D_L}{\rho}.
\end{align}
Based on the noise power spectral density $S_h(f)$ (PSD) of ET given
in \citet{Zhao11}, the Signal-to-Noise ratio (SNR) of the network of
three independent interferometers can be calculated as
\begin{equation}
\rho=\sqrt{\sum\limits_{i=1}^{3}\left\langle
\mathcal{H}^{(i)},\mathcal{H}^{(i)}\right\rangle}. \label{euqa:rho}
\end{equation}
Here the inner product is defined as
\begin{equation}
\left\langle{a,b}\right\rangle=4\int_{f_{\rm lower}}^{f_{\rm
upper}}\frac{\tilde a(f)\tilde b^\ast(f)+\tilde a^\ast(f)\tilde
b(f)}{2}\frac{df}{S_h(f)}, \label{euqa:product}
\end{equation}
where the lower limit in frequency of ET is $f_{\rm lower}=1$Hz, the
upper cutoff frequency is set as $f_{\rm upper}=2/(6^{3/2}2\pi
M_{\rm obs})$ and $M_{\rm obs}=(1+z)M_{\rm phys}$ is the observed
total mass. Note that a GW detection is confirmed if a network SNR
is $\rho>8.0$, the SNR threshold currently used by LIGO/Virgo
network \citep{Cai17}. The second source of uncertainty is generated
by gravitational lensing, which is always modeled as
$\sigma_{D_L}^{lens}/D_L=0.05z$ \citep{Sathyaprakash10,Li15,Cai17}.
Thus, the total uncertainty of $D_L$ is taken to be
\begin{align}
\sigma_{D_L}&~~=\sqrt{(\sigma_{D_L}^{\rm inst})^2+(\sigma_{D_L}^{\rm lens})^2} \nonumber\\
            &~~=\sqrt{\left(\frac{2D_L}{\rho}\right)^2+(0.05z D_L)^2}.
\label{sigmadl}
\end{align}
Thus, the total uncertainty on luminosity distance is
\begin{align}
\sigma_{D_L}^2=(\sigma_{D_L}^{\rm inst})^2+(\sigma_{D_L}^{\rm
lens})^2.
\end{align}

\begin{figure}
\centering
\includegraphics[width=9cm,height=8cm]{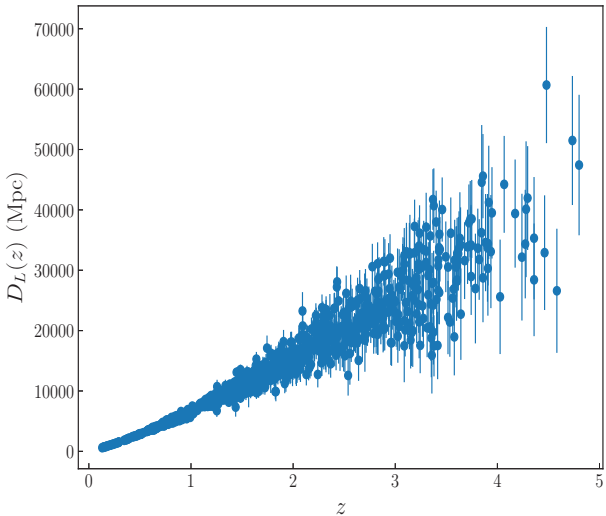}
\caption{The 1000 simulated GW events from the third-generation
gravitational wave detector, i.e., the Einstein
Telescope.}\label{DL}
\end{figure}

Focusing on the expected rates of BNS and BHNS detections per year
for the ET, only a small fraction of binary mergers may have the
observation of detectable EM counterparts like short and intense
bursts of $\gamma$-rays (SGRB). In this work we follow the recent
analysis of \citet{Cai17} and assume that ET can detect up to 1000
GW events in a 10 year observation (with measurable source
redshift). As for the redshift distribution of the observable GW
sources, we follow the function taking into account evolution and
stellar synthesis \citep{Schneider01,Cutler09,Sathyaprakash10},
which has been widely applied in gravitational wave cosmology
\citep{Qi19}. The fiducial cosmological model we choose is the
concordant model based on the most recent Plank results
\citep{Ade16}. The GW mock data covering the redshift range $z\sim
5$ are shown in Fig.~\ref{DL}, with the luminosity distance
measurements obeying the Gaussian distribution
$D_L^{mean}=\mathcal{N}(D_L^{\rm fid},\sigma_{D_L})$.

In this analysis, we also include the latest observational data
derived in the electromagnetic (EM) window, including the recently
released Pantheon type Ia supernovae sample, the baryon acoustic
oscillation (BAO) observations, and Hubble parameter
measurements obtained through the cosmological-model-independent
method using the so-called cosmic chronometers
\citep{Jimenez2002,Jimenez2003}. The type Ia supernovae sample used
in our analysis include 1048 SNe Ia observed by the Pan-STARRS1
(PS1) Medium Deep Survey, covering the redshift range of
$0.01<z<2.3$ \citep{Scolnic18}. For the BAO data, we use the six
measurement of distance ratio $\frac{d_A(z*)}{D_V(z_{BAO})}$ in the
redshift range $0.106\leq z\leq0.73$, from the 6-degree Field Galaxy
Survey (6dFGS), the WiggleZ galaxy survey, and the Sloan Digital Sky
Survey (SDSS-DR7) catalog combined with galaxies from 2dFGRS
\citep{Beutler11,Ross15,Alam17,Zhao18,Des17}. Here $D_V$ is the
dilation scale
\begin{equation}
D_V(z)=\left(d_A(z)^2\frac{z}{H(z)}\right)^{1/3},
\end{equation}
and $d_A$ is the co-moving angular diameter distance \citep{Xu18}
\begin{equation}
d_A(z)=\int^z_0\frac{dz'}{H(z')}.
\end{equation}
The specific BAO data and their corresponding fitting procedures are
summarized in the source papers
\citep{Simon2005,Stern2010,Moresco2012,Moresco2015,Moresco2016,Zhang14}.
For the Hubble parameter data, 30 measurements of $H(z)$
covering the redshift range $0.07\leq z \leq 1.965$ have been
obtained, based on the differential age evolution of passively
evolving early-type galaxies \citep{Moresco2015,Moresco2016}. The
possible cosmological application of these cosmic chronometers has
been extensively discussed in the literature
\citep{Ratsimbazafy2017}, such as the reconstruction of cosmological
distances \citep{Wu19,Zheng21} and the spatial curvature of the
universe \citep{Liu20b} in a cosmological model-independent way. In
particular, \citet{Nunes17} have recently studied the advantage of
utilizing the Hubble parameter data to impose improved constraints
on the viable and most used $f(R)$ gravity models, which encourages
us to improve and develop it further in this analysis.

\begin{figure*}
\centering
\includegraphics[width=10cm,height=10cm]{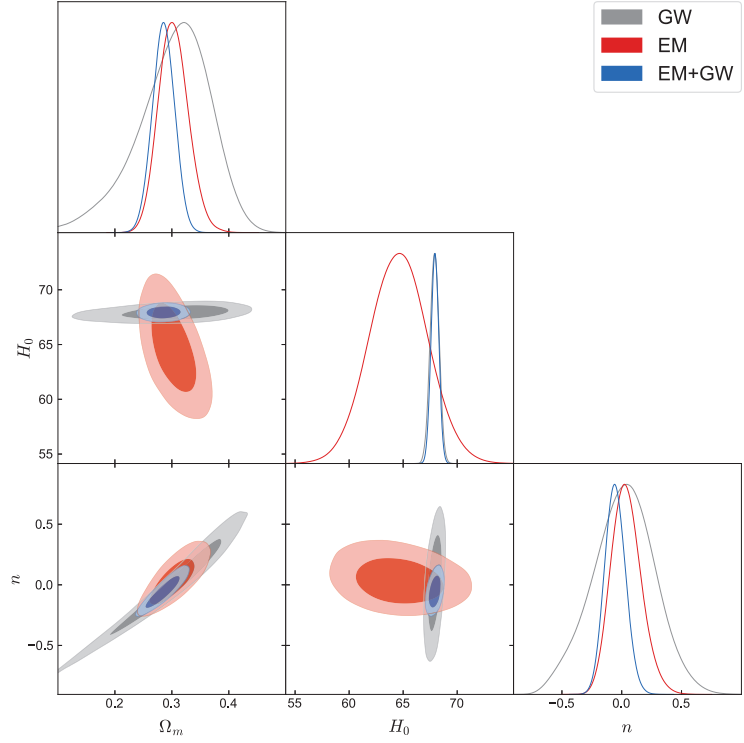}
\caption{The likelihood distributions of $\Omega_m$, $n$ and
$H_0$ in the $f_1(\mathcal{R})$ model, from the simulated GW data
and its combination with the current EM observations.}\label{f1gw}
\end{figure*}

\begin{figure*}
\centering
\includegraphics[width=10cm,height=10cm]{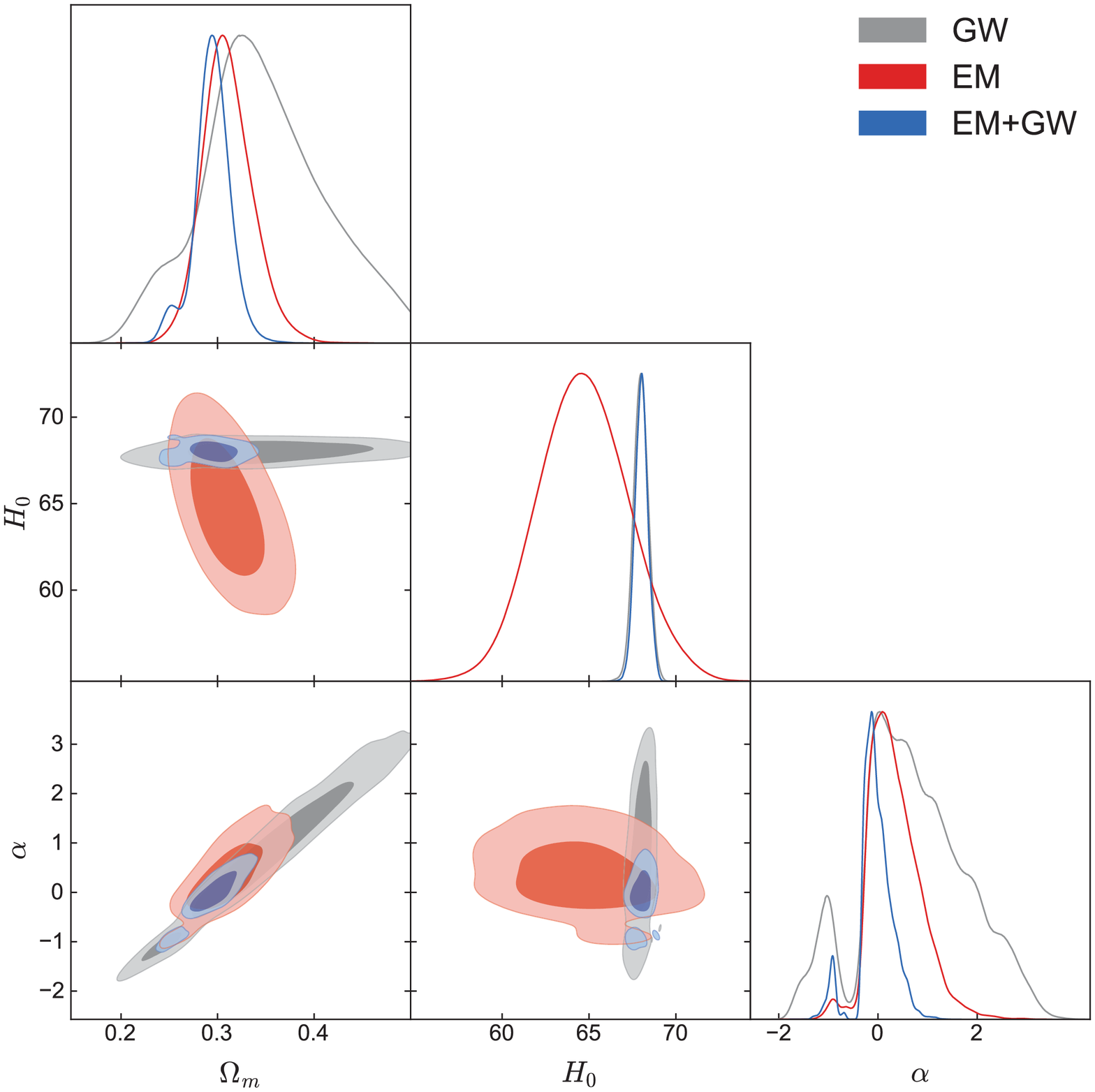}
\caption{The likelihood distributions of $\Omega_m$,
$\alpha$ and $H_0$ in the $f_2(\mathcal{R})$ model, from the
simulated GW data and its combination with the current EM
observations.}\label{f2gw}
\end{figure*}

\begin{table*}
 \begin{center}{\scriptsize
 \begin{tabular}{|c|c|c|c|} \hline\hline
 \cline{1-4}
 Data  \ \ & \ \ $\Omega_{m}$   &$H_0$   &$n$ \\ \hline
 EM+GW\ \ & \ \ $0.285\pm0.019(1\sigma)\pm{0.038}\sigma)$ \ \ & \ \ $67.94\pm0.36(1\sigma)\pm0.70(2\sigma)$\ \ & \ \ $-0.055\pm0.086(1\sigma)^{+0.17}_{-0.16}(2\sigma)$\ \\ \hline
 EM\ \  & \ \ $0.303^{+0.024}_{-0.028}(1\sigma)^{+0.054}_{-0.049}(2\sigma)$ \ \ & \ \ $64.6\pm2.7(1\sigma)^{+5.3}_{-5.2}(2\sigma)$ \ \ & \ \ $0.04^{+0.11}_{-0.14}(1\sigma)^{+0.25}_{-0.24}(2\sigma)$\ \\ \hline
 GW\ \ & \ \ $0.303^{+0.073}_{-0.050}(1\sigma)^{+0.12}_{-0.13}(2\sigma)$ \ \ & \ \ $67.91\pm0.42(1\sigma)\pm0.82(2\sigma)$ \ \   & \ \ $0.01^{+0.27}_{-0.24}(1\sigma)^{+0.50}_{-0.53}(2\sigma)$\ \\
 \hline \hline
Data  \ \ & \ \ $\Omega_{m}$   &$H_0$   &$\alpha$ \\ \hline
 EM+GW\ \ & \ \ $0.293^{+0.017}_{-0.015}(1\sigma)^{+0.035}_{-0.046}(2\sigma)$ \ \ & \ \ $68.03^{+0.29}_{-0.33}(1\sigma)^{+0.69}_{-0.62}(2\sigma)$\ \ & \ \ $-0.06\pm0.35(1\sigma)^{+0.70}_{-0.97}(2\sigma)$\ \\ \hline
 EM\ \  & \ \ $0.310^{+0.023}_{-0.029}(1\sigma)^{+0.056}_{-0.048}(2\sigma)$ \ \ & \ \ $64.6\pm2.6(1\sigma)^{+5.2}_{-5.1}(2\sigma)$ \ \ & \ \ $0.35^{+0.31}_{-0.59}(1\sigma)^{+1.2}_{-0.93}(2\sigma)$\ \\ \hline
 GW\ \ & \ \ $0.347\pm0.063(1\sigma)^{+0.13}_{-0.12}(2\sigma)$ \ \ & \ \ $67.99\pm0.42(1\sigma)^{+0.83}_{-0.81}(2\sigma)$ \ \   & \ \ $0.7\pm1.1(1\sigma)^{+2.1}_{-2.0}(2\sigma)$\ \\
 \hline\hline
 \end{tabular}}
 \end{center}
\caption{The marginalized $1\sigma$ uncertainties of the
parameters $\Omega_{m}$, $H_0$ and $n$ ($\alpha$) for the two viable
models in $f(R)$ theories.} \label{tablefR1}
 \end{table*}

 \begin{figure*}
\centering
\includegraphics[width=8cm,height=6cm]{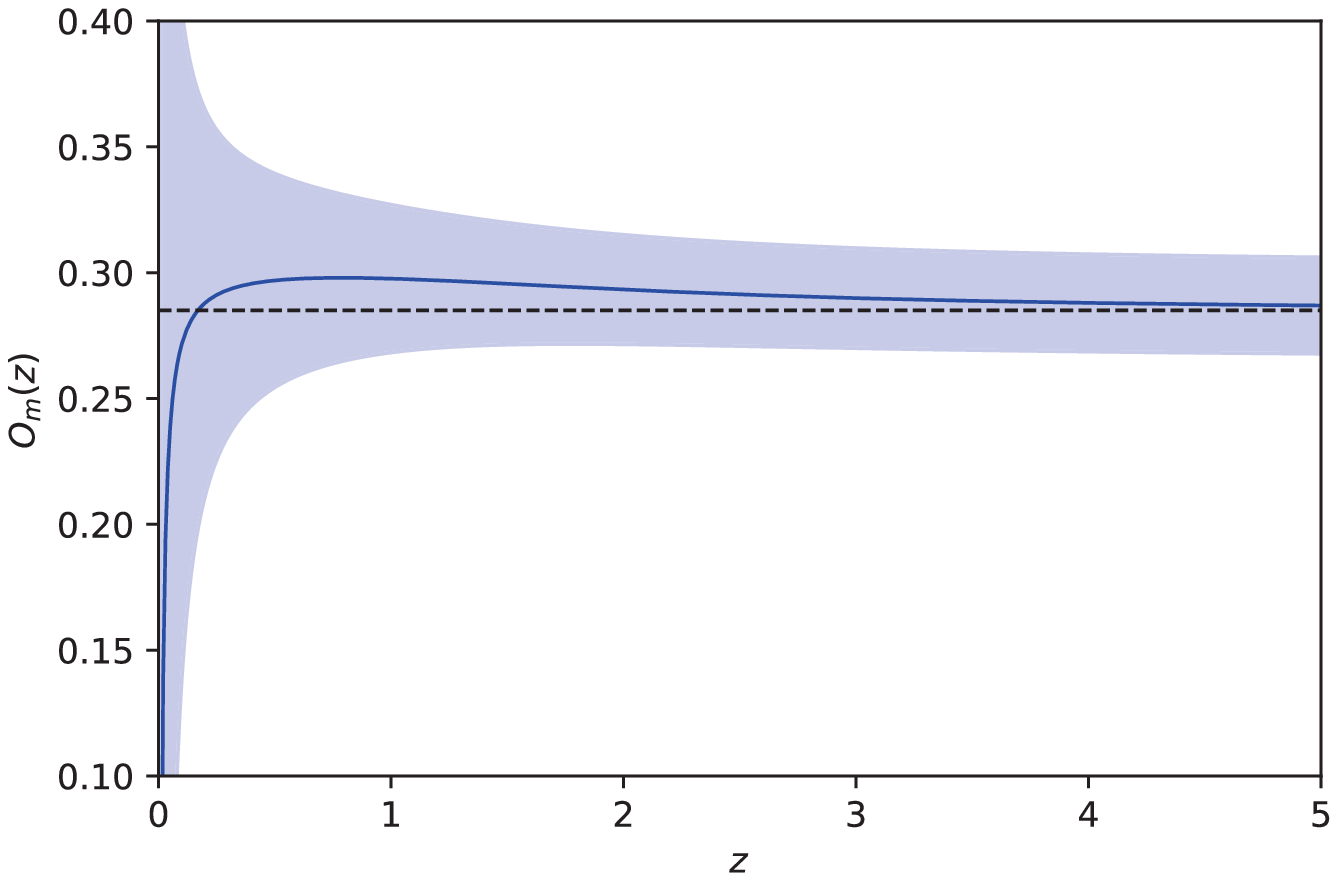}\includegraphics[width=8cm,height=6cm]{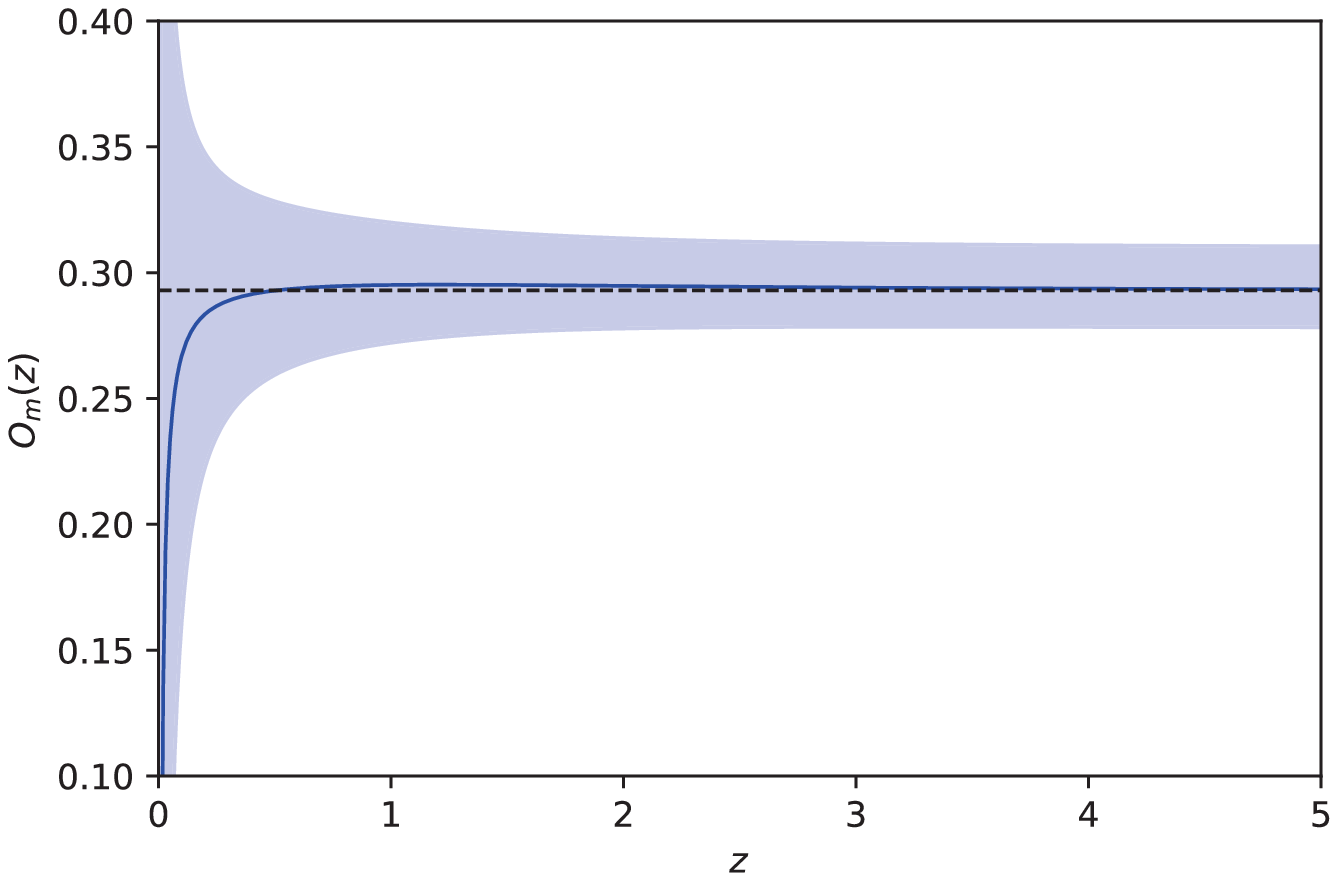}
\caption{The evolution of $O_m(z)$ versus the redshift $z$
for $f_1(\mathcal{R})$ and $f_2(\mathcal{R})$ model from the
simulated EM+GW data, with the $1\sigma$ uncertainty denoted by blue
shades. The black dashed lines represent the standard $\Lambda$CDM
model.}\label{fgw}
\end{figure*}

\begin{figure*}
\centering
\includegraphics[width=8cm,height=6cm]{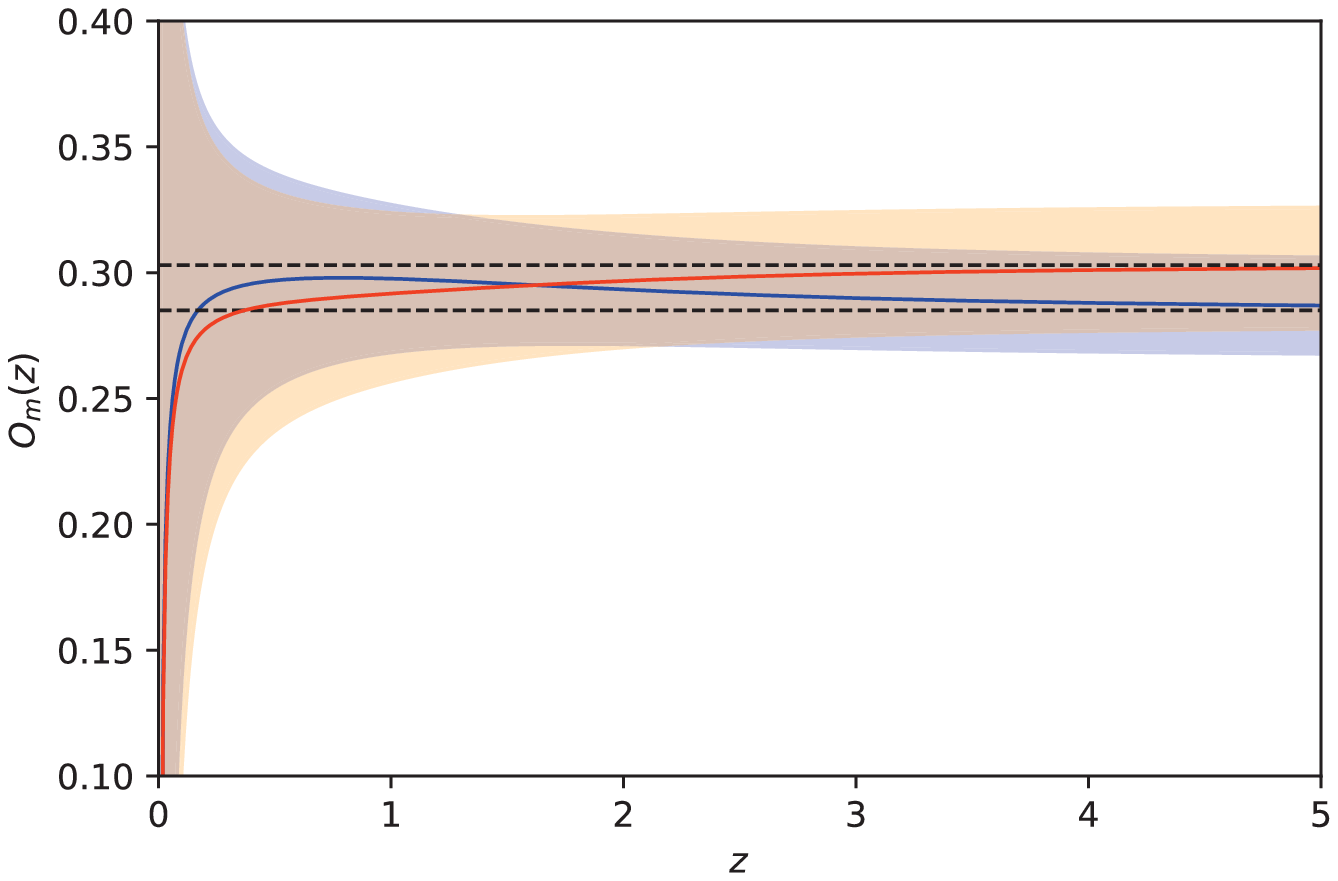}\includegraphics[width=8cm,height=6cm]{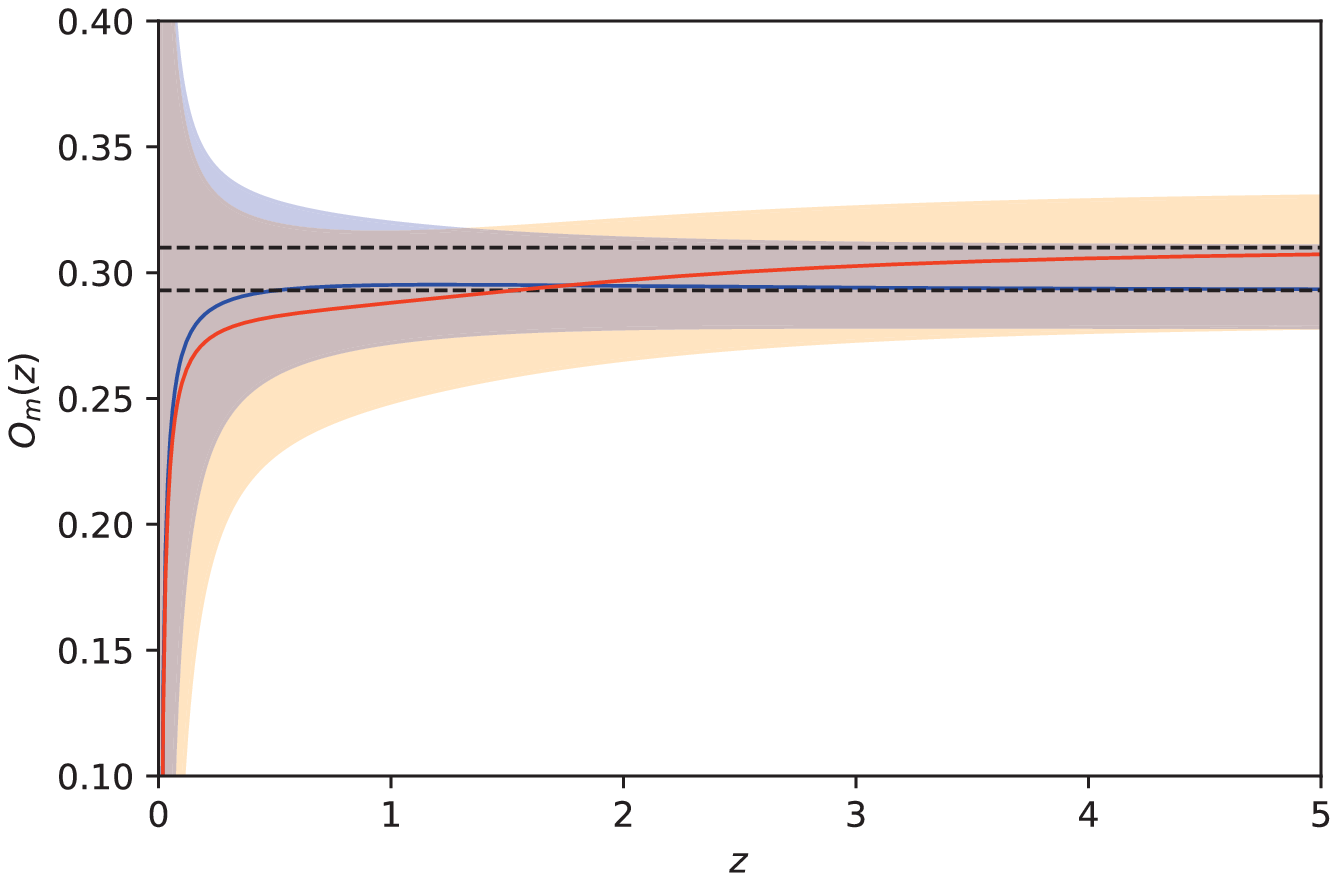}
\caption{The evolution of $O_m(z)$ versus the redshift $z$
for $f_1(\mathcal{R})$ and $f_2(\mathcal{R})$ model from the EM+GW
data and EM data, with the $1\sigma$ uncertainty denoted by blue
shades(EM+GW), with the $1\sigma$ uncertainty denoted by oronge
shades(EM), The black dashed lines represent the standard
$\Lambda$CDM model.}\label{fsngw}
\end{figure*}

\section{Constraints and model diagnostics}\label{result}

\subsection{Observational constraints on the $f(R)$ models}

In this section we will investigate observational bounds on two
$f(R)$ models from a statistical analysis involving three classes of
data combinations, i.e., GW, EM (SNe+BAO+$H(z)$), and GW+EM
(GW+SNe+BAO+$H(z)$). The Monte Carlo Markov Chain (MCMC) method is
applied to perform the minimum likelihood method of $\chi^{2}$ fit
\citep{Lewis02}, with the fitting results shown in Table 1. The
marginalized probability distribution of each parameter ($\Omega_m$,
$n$ and $H_{0}$ for $f_1(\mathcal{R})$; $\Omega_m$, $\alpha$ and
$H_{0}$ for $f_2(\mathcal{R})$) and the marginalized 2D confidence
contours are presented in Fig.~2-3.

Let us start from the first $f(R)$ theory within the Palatini
formalism:
$f_{1}(\mathcal{R})=\mathcal{R}-\frac{\beta}{\mathcal{R}^{n}}$. For
the simulated GW data, the best-fit values of the parameters in the
$f_1(\mathcal{R})$ model are $\Omega_{m}=0.303^{+0.073}_{-0.050}$,
$H_0=67.91\pm0.42$ km/s/Mpc, and $n=0.01^{+0.27}_{-0.24}$. In order
to have a good comparison, the fitting results derived from the EM
observations (SNe+BAO+$H(z)$) are also shown in Fig.~2, with the
best-fit parameters of $\Omega_m=0.303^{+0.024}_{-0.028}$,
$H_0=64.6\pm2.7$km/s/Mpc, and $n=0.04^{+0.11}_{-0.14}$. One should
note that, although GWs alone seem not able to provide the
constraints as good as the current EM observations (SNe+BAO+$H(z)$),
the degeneracy between different $f(\mathcal{R})$ parameters
obtained from the standard sirens is different from the statistical
standard probes in the EM window. Such tendency could also be
clearly seen from the comparison between the plots obtained with the
joint GW+EM analysis, i.e., with the combined standard probe data
sets in the GW and EM data sets, the best-fit value for the
parameters are $\Omega_{m}=0.285\pm0.019$,
$H_0=67.94\pm0.36$km/s/Mpc, and $-0.055\pm0.086$. Therefore, the
shift in the best-fitted parameters (with remarkably reduced allowed
region) illustrates how the combination of the future GW
observations can be used to improve the model parameters in
$f_1(\mathcal{R})$ cosmology. This has been noted by the previous
analysis studying the constraint ability of the gravitational wave
(GW) as the standard siren on the cosmological parameters
\citep{Cai17}.

The best-fit values of the parameters along with their $1\sigma$
uncertainties from three different data combinations are explicitly
presented in Table \ref{tablefR1}. Note also that in the modified
$f_1(\mathcal{R})$ gravity, the best-fit value for the matter
density parameter and the Hubble parameter, i.e., $\Omega_{m} =
0.285\pm0.019$ and $H_0=67.94\pm0.36$km/s/Mpc, are well consistent
with the current estimates in the framework of $\Lambda$CDM
cosmology given by recent Planck data release \citep{Aghanim18}.
More importantly, the deviation parameter $n$ from three
combinations of data analysis (EM+GW, EM and GW) indicates that
$\Lambda$CDM model ($n=0$) is still included within 68.3\%
confidence level. However, we can see that the best value of
$n=-0.055$ seems to be slightly smaller than $0$, which suggests
that there still exists some possibility that $\Lambda$CDM may not
be the best cosmological model preferred by the current
observations. Such conclusion is marginally different from the
previous results obtained in \citet{Santos08,Amarzguioui06,Fay07},
using the supernova Gold and the SNLS data sets respectively.

Next we turn to study the ability of the standard siren to infer the
parameters in the $f_2(\mathcal{R})$ model:
$f_{2}(\mathcal{R})=\mathcal{R}+\alpha\ln{\mathcal{R}}-\beta$. The
best fits takes place at $\Omega_{m}=0.347\pm0.063$,
$H_0=67.99\pm0.42$ km/s/Mpc and $\alpha=0.7\pm1.1$ with the
simulated GW data. The constraining power of the standard sirens in
breaking degeneracy between model parameters is slightly existence, as can
be seen from the marginalized 1$\sigma$ and 2$\sigma$ contours of
each parameter in Fig.~3. By fitting the $f_{2}(\mathcal{R})$ model
to the GW+EM observations, we obtain $\Omega_{m}=0.293^{+0.017}_{-0.015}$,
$H_0=68.03^{+0.29}_{-0.33}$ km/s/Mpc and $\alpha=-0.06\pm0.35$. In Table 1 we
also summarize the best-fit values for the three combined data sets,
respectively. In broad terms, the estimated values of $\Omega_m$ and
$H_0$ are basically consistent with each other at the 68.3\%
confidence level. Moreover, compared with the case in the
$f_1(\mathcal{R})$ model, the largest difference happens on the
constraint of $\alpha$: the deviation from $\Lambda$CDM also tends
to be slightly smaller than $0$, although the concordance
cosmological scenario is still included within 68.3\% confidence
level. Therefore, there still exists some possibility that
$\Lambda$CDM may not the best cosmological model preferred by the
current and future high-redshift GW+EM observations. The
constraining ability of standard siren data can be quantified by
comparing the results at 1$\sigma$ C. L. based on GW+EM analysis,
i.e., we obtain the error bar smaller than that from EM data
alone, when the $f_2(\mathcal{R})$ model is considered.

\begin{figure*}
\centering
\includegraphics[width=8cm,height=6cm]{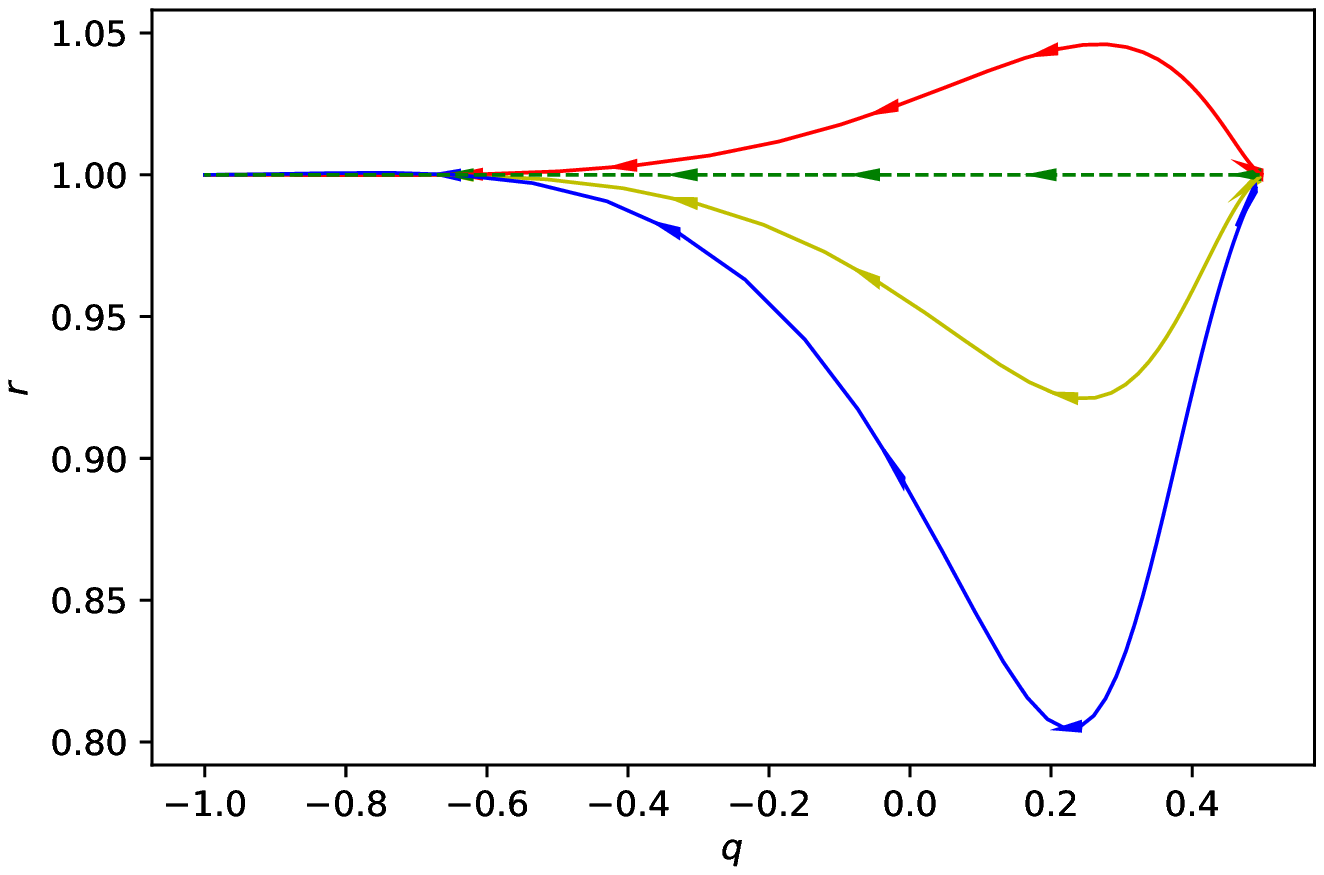}\includegraphics[width=8cm,height=6cm]{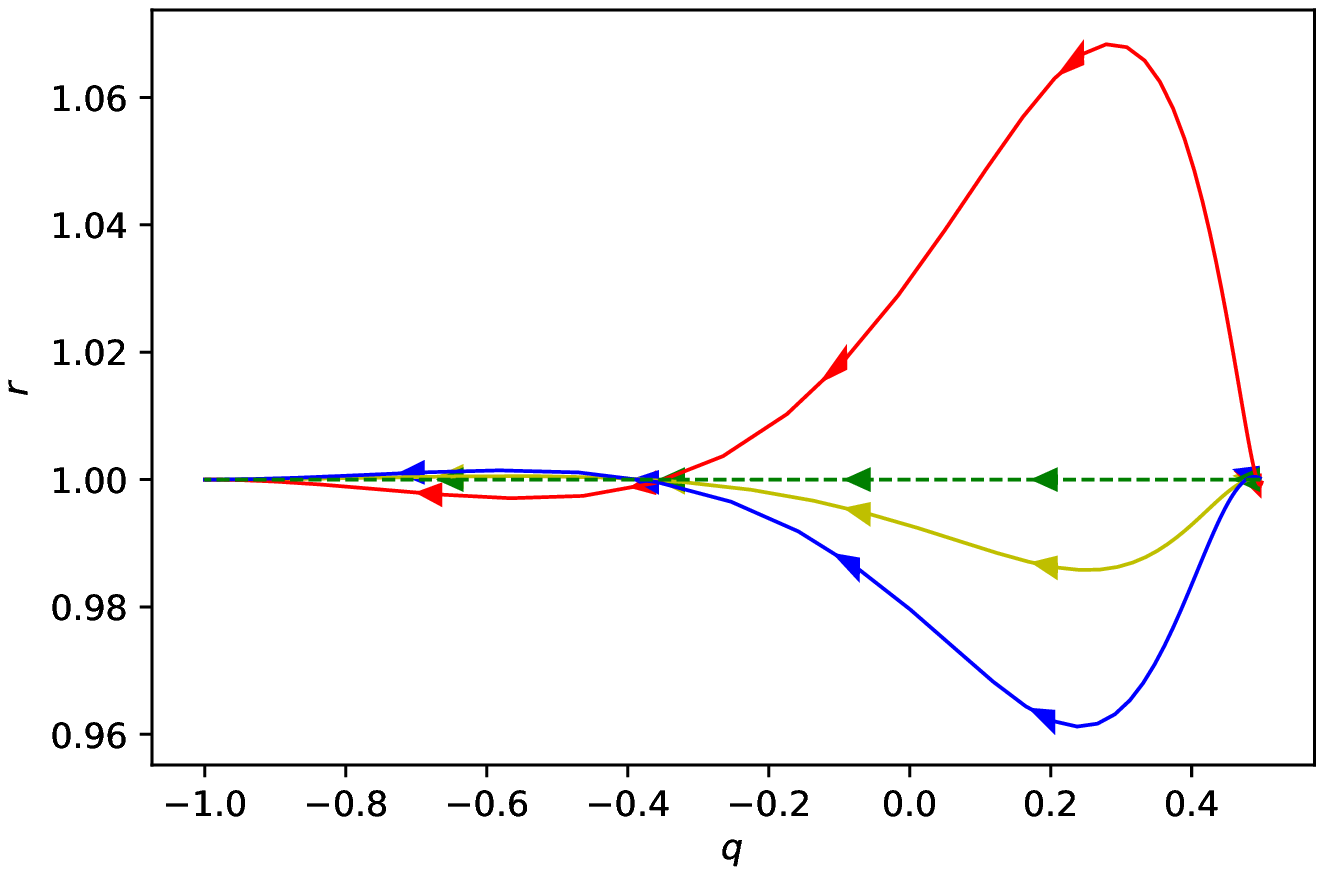}
\caption{The plot shows the behavior of the
$f_1(\mathcal{R})$ model and $f_2(\mathcal{R})$ model in the $(r,q)$
plane. The red and blue lines represent the evolution of $(r,q)$
within the 1$\sigma$ error range.}\label{frrq}
\end{figure*}

\begin{figure*}
\centering
\includegraphics[width=8cm,height=6cm]{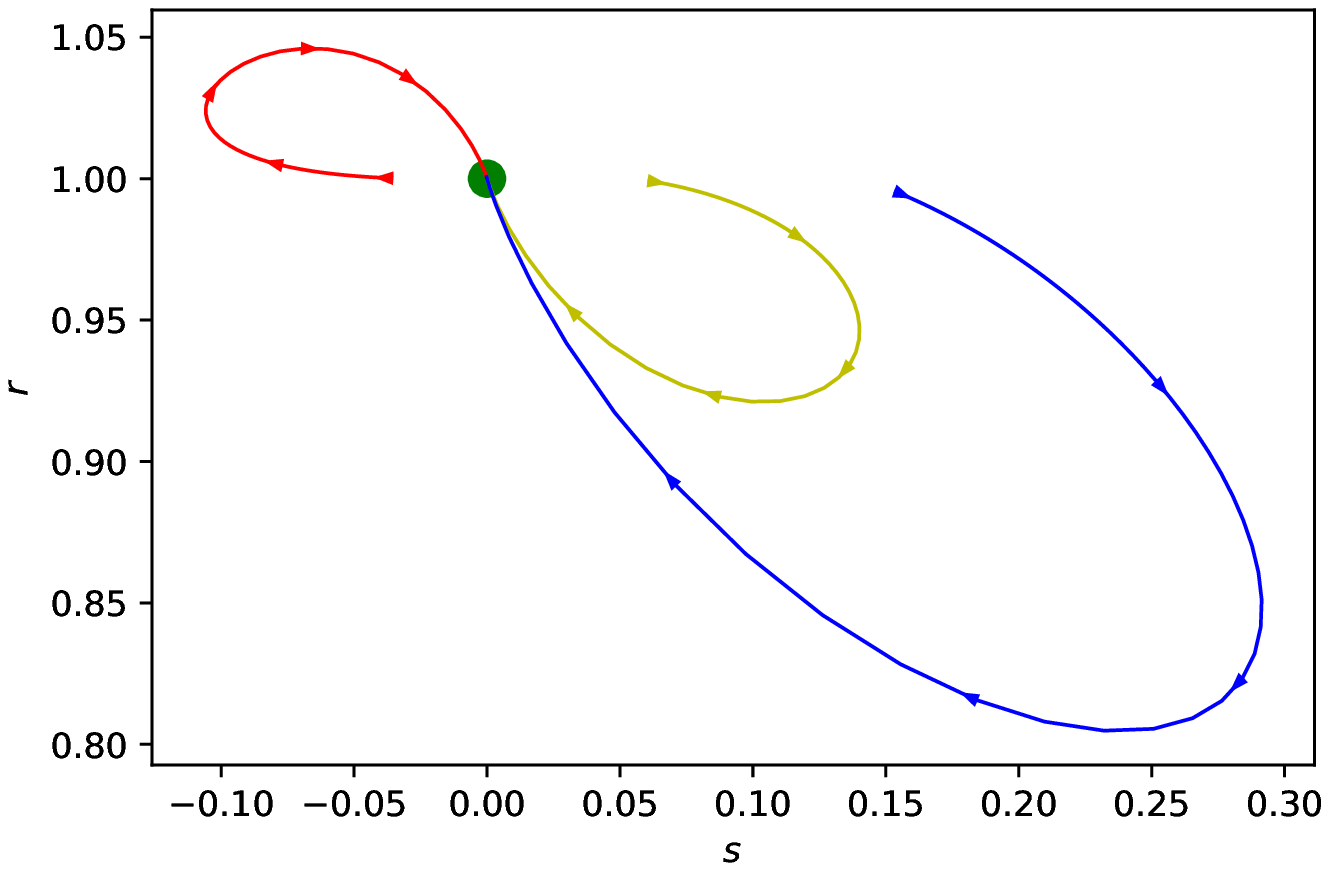}\includegraphics[width=8cm,height=6cm]{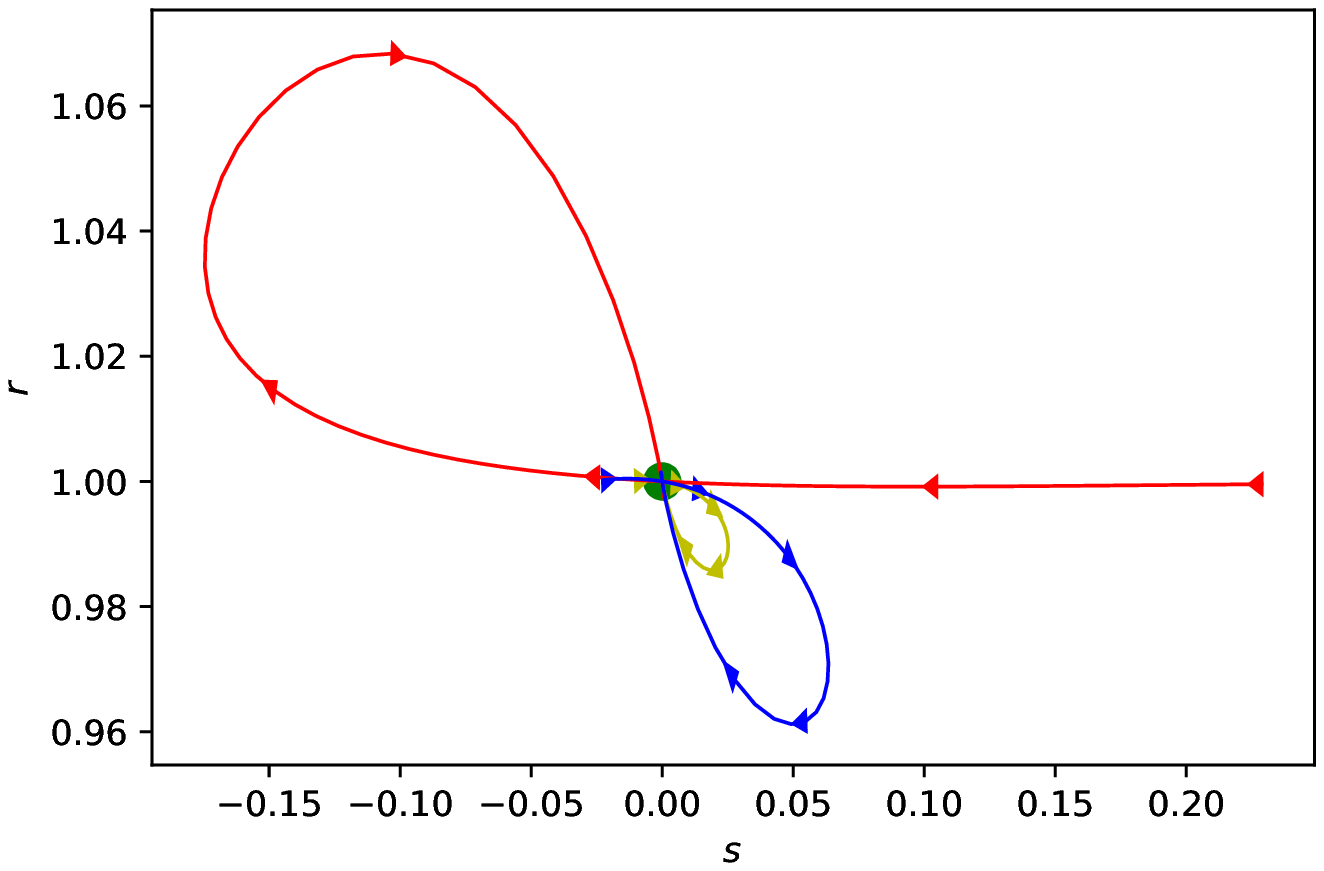}
\caption{The same as Fig.~6, but in the $(r, s)$
plane.}\label{frrs}
\end{figure*}

\subsection{Model diagnostics}

Now two diagnostics analysis (the $Om(z)$ and the statefinder
diagnostics) will be performed in this section, acting as
alternative probes capable of discriminating between $f(R)$
cosmology and $\Lambda$CDM model.

On the one hand, the expansion rates at different redshifts, or the
Hubble parameters $H(z)$, has opened a new chapter of using the
so-called $Om(z)$ diagnostics to discriminate different cosmological
models as well as $\Lambda$CDM model
\citep{Sahni08,Ding15,Qi18,Zheng16,Zheng19}. In this method, only a
single time derivative of $a(t)$ is used and a new diagnostics is
defined as
\begin{equation}
Om(z)=\frac{E^2(z)-1}{(1+z)^3-1}
\end{equation}
where $E(z)=\frac{H(z)}{H_0}$ represents the dimensionless expansion
rate. It is obvious that in the flat $\Lambda$CDM model, the $Om(z)$
evaluated at any redshift is always equal to the present mass
density parameter $\Omega_m$. Therefore, such diagnostic can be used
as a cosmological probe to directly illustrate the difference
between $\Lambda$CDM and other cosmological models.

Based on the best-fit model parameters and their $1\sigma$
uncertainties derived from the simulated EM+GW data, we have different
values of $Om(z)$ as the function of redshift for the $\Lambda$CDM
and two $f(R)$ models, which are presented in Fig.~\ref{fgw}. It is
obvious that the $O_m(z)$ curve of the $f_1(\mathcal{R})$ model will
coincide with $\Lambda$CDM at high redshifts ($z>4$), which
indicates that the $O_m(z)$ for the $f_1(\mathcal{R})$ model cannot
be distinguished $\Lambda$CDM in the early universe. However, the
$f_1(\mathcal{R})$ cosmology begins to deviate from $\Lambda$CDM in
the redshift range of $z<4$, while the deviation between
$f_2(\mathcal{R})$ model and $\Lambda$CDM takes place at the
redshift of $z>1.5$. In Fig.~\ref{fsngw} we also compare the
$O_m(z)$ function based on the fits from EM+GW and EM data. One may
clearly see that in the framework of two $f(R)$ models, the
combination of future naturally improved GW data most recent EM
observations will reveal the consistency or acknowledge the tension
between the $\Lambda$CDM model and modified gravity theories.

On the other hand, in order to more effectively differentiate and
compare between different cosmological models, we turn to a
discussion on the so-called statefinder diagnostic. For a specific
cosmological model, the Hubble parameter $H$ and the deceleration
parameter $q$ are respectively related as
\begin{equation}\label{hq}
q(z)=-\frac{\ddot{a}}{aH^2}=\frac{E'(z)}{E(z)}(1+z)-1
\end{equation}
In the analysis of statefinde diagnostic \citep{Sahni08}, the new
dimensionless parameters ${r,s}$ are constructed from the scale
factor $a(t)$ and its time derivatives similar to the geometrical
parameters $H(z)$ and $q(z)$:
\begin{eqnarray}
r(z)&=&q(z)(1+2q(z))+q'(z)(1+z) \\ \nonumber
s(z)&=&\frac{1}{3}(r(z)-1)/(q(z)-1/2)
\end{eqnarray}
where $q'(z)\equiv\frac{dq(z)}{dz}$. Any obtained model can be
compared with the standard cosmological constant model, from which
one may see how a model approaches or deviates from $\Lambda$CDM in
the $s-r$ plane and $q-r$ plane. The evolution of starefinder pair
$(s,r)$ in the two $f(R)$ models is shown in Fig.~\ref{frrs}. On the
one hand, it is obvious that the statefinder pair exhibits the
similar behaviors in the framework of two $f(R)$ models, i.e.,
originating from the same point of $(r,q)=(1,0.5)$, evolving along
different trajectory, and finally converging on the same point of
($1,-1$). On the other hand, compared with the deceleration
parameter $q$, the parameter $r$ exhibiting apparent fluctuations is
more effective in discriminating different cosmological models. The
trajectories of the two $f(R)$ models in the $r-q$ plane are also
presented in Fig.~\ref{frrq}, in which the point $(r,s)=(1,0)$
corresponding to the $\Lambda$CDM universe is also added for
comparison. It is noteworthy that, although the corresponding value
for the two $f(R)$ models significantly deviates from $\Lambda$CDM
at the present epoch, both of  two modified gravity models discussed
in this analysis is resembling to the standard cosmological model in
the future and ultimately freezing to it. Such a tendency is more
obvious when the $1\sigma$ uncertainties of the $f(R)$ parameters
are taken into consideration.

\section{Conclusions} \label{conclusion}

In this paper we have investigated the constraint ability of GW
events on the modified gravity models using the simulated data from
the third-generation gravitational wave detector, the Einstein
Telescope. We focus on two $f(R)$ models which introduces a
perturbation of the Ricci scalar $R$ in the Einstein-Hilbert action,
which can be considered as an important candidate models alternative
to dark energy. More specifically, in the framework of two viable
models in $f(R)$ theories within the Palatini formalism
($f_{1}(\mathcal{R})=\mathcal{R}-\frac{\beta}{\mathcal{R}^{n}}$ and
$f_{2}(\mathcal{R})=\mathcal{R}+\alpha\ln{\mathcal{R}}-\beta$), our
results show that the sensitivity achieved by the ET detector or a
similar third generation interferometer is enough to improve the
current estimates on the free parameter within $f(R)$ gravity. In
practise more GW events may be needed in order to achieve the
accuracy of EM data. More importantly, the standard sirens GWs could
effectively eliminate the degeneracies among parameters in the two
modified gravity models, in joint analysis with the latest
electromagnetic (EM) observational data (including the recently
released Pantheon type Ia supernovae sample, Hubble
parameter data from cosmic chronometers, and baryon acoustic
oscillation distance measurements).

In addition, we thoroughly investigate the nature of geometrical
dark energy in the modified gravity theories with the assistance of
$Om(z)$ and statefinder diagnostic analysis. The present analysis
makes it clear that the simplest cosmological constant model is
still the most preferred by the current data. However, the
combination of future naturally improved GW data most recent EM
observations will reveal the consistency or acknowledge the tension
between the $\Lambda$CDM model and modified gravity theories.
Therefore, in the case of detections with the ET, our forecast
analysis indicates that it is possible to provide an independent and
complementary alternative to current experiments, once that these
events begin to be detected. One should note that, dedicated
observations of the sky position of each host galaxy (i.e., with
negligible measurement error) would be necessary, in order to obtain
significant improvements in the distance, and hence in the
constraints on modified gravity theories. Although the GW distance
posterior changes slowly over the sky and therefore is not sensitive
to the precise location of the counterpart, obtaining such
measurements for a sample of different types of GW events would
require substantial follow-up efforts \citep{Zhang20}.

\section*{Acknowledgments}

We thank Prof. Z.-H. Zhu for useful discussions. This work was
supported by National Key R\&D Program of China No. 2017YFA0402600;
the National Natural Science Foundation of China under Grants Nos.
12021003, 11690023, and 11633001; Chongqing Municipal Science and
Technology Commission Fund (cstc2018jcyjAX0192); Beijing Talents
Fund of Organization Department of Beijing Municipal Committee of
the CPC; the Strategic Priority Research Program of the Chinese
Academy of Sciences, Grant No. XDB23000000; the Interdiscipline
Research Funds of Beijing Normal University; and the Opening Project
of Key Laboratory of Computational Astrophysics, National
Astronomical Observatories, Chinese Academy of Sciences. J. Wang was
supported by the National Natural Science Foundation of China under
Grant No.11647079, Science and Technology Department of Yunnan
Province - Yunnan University Joint Funding (2019FY003005), Donglu
Youth Teacher Plan of Yunnan University and the Key Laboratory of
Astroparticle Physics of Yunnan Province.

\end{document}